\documentstyle[seceq,epsbox,preprint]{jpsj}

\def\runtitle
{Analysis on Aging in the Generalized Random Energy Model}
\def\runauthor{Munetaka {\sc Sasaki} and Koji {\sc Nemoto}}
\title{Analysis on Aging in the Generalized Random Energy Model}
\author{Munetaka {\sc Sasaki} and Koji {\sc Nemoto}}

\inst{Division of Physics, Hokkaido University, Sapporo 060-0810}
\recdate{\phantom{February 3, 2000}}
\abst{
A new dynamics more natural than that proposed by Bouchaud and Dean 
is introduced to the Generalized Random Energy Model, and 
the master equation for the dynamics is 
solved exactly to calculate the time correlation function 
\( C(t+t_{\rm w},t_{\rm w}) \). Although our results are very similar 
to those obtained by Bouchaud and Dean qualitatively, the exponents for 
power law relaxation are different. 
The Zero-Field-Cooled magnetization is also 
calculated with a relation between the correlation function and the response 
function which holds even if the relaxation is non-equilibrium. The validity 
of these analytic results are confirmed by numerical simulations. 
}
\kword{aging, generalized random energy model, zero-field-cooled magnetization
}
\begin{document}
\sloppy
\maketitle
\section{Introduction}\label{sec:introduction}
The aging, dynamical behavior largely depending on the 
history of system, is one of the most striking phenomena in the complex 
systems such as spin glasses, glasses, polymers and proteins, and 
many theoretical models have been proposed to explain these phenomena. 
The Generalized Random Energy Model (GREM)\cite{GREM2,GREM3,Bouchaud} 
is one of such models. 
Bouchaud and Dean calculated the correlation function 
\( C(t+t_{\rm w},t_{\rm w}) \) exactly and showed that 
\( t/t_{\rm w}\)-scaling holds in this model\cite{Bouchaud}, 
although the dynamics they introduced seems to be a little unnatural. 
We recently introduce more natural dynamics to this model, 
and carried out simulations\cite{condmat} on aging phenomena 
with temperature variations. As a result, we revealed that 
the fundamental features reported in these complex systems, 
such as the memory effect and the rejuvenation, were well reproduced along the 
hierarchical picture discussed in ref.~\citen{HP1,HP2,HP3}. 

In this paper, we calculate the exact correlation function 
for dynamics used in ref.~\citen{condmat}, 
and we confirm that while the \( t/t_{\rm w}\)-scaling 
holds as ref.~\citen{Bouchaud}, our result is different from theirs 
on exponents for power law relaxation. By introducing magnetizations 
to this model\cite{condmat}, 
we also calculate the ZFC magnetization 
\( M_{\rm ZFC}(t,t_{\rm w}) \)\cite{ZFCexperiment1,ZFCexperiment2}, 
which is measured by quenching the system from above \( T_{\rm c} \) 
to below and applying a weak magnetic 
field during \( t \) after a waiting time \( t_{\rm w} \). 
We estimate \( M_{\rm ZFC}(t,t_{\rm w}) \) by using a relation 
between the response 
function \( R(t,t') \) and the correlation function 
\( C(t,t') \)\cite{hierarchical1}
\begin{equation}
 R(t,t')=-\frac1T\frac{\rm d}{{\rm d} t} C(t,t'),
\label{eqn:relationRC}
\end{equation}
where \( T \) is the temperature. This relation is valid when assuming that 
the external field \( H \) is small enough and the transition 
probability from a state \( \beta \) to \( \alpha \) is modified 
by a factor \( \exp(-\frac{H M_{\beta}}{T}) \), where \( M_{\beta} \) is 
the magnetization of a state \( \beta \). This condition is 
satisfied in the present way of introducing magnetizations\cite{condmat}. 
Although this relation is very similar to the Fluctuation 
Dissipation Theorem (FDT) for equilibrium relaxation 
(actually it reduces to FDT if time homogeneity holds), it is valid even if 
the relaxation is non-equilibrium. 
These analytic results on the correlation function and 
the ZFC magnetization are checked by numerical simulations.

The organization of this manuscript is following. In section 2, we
introduce the GREM. In section 3, we calculate the correlation function 
and the ZFC magnetization analytically and check these results 
by simulations. In section 4, we summarize this paper with some discussions. 
\section{Model}
The GREM is schematically shown in Fig.~1. The bottom points 
represent the accessible states of this system. This model is constituted by 
piling up \( L \) layers hierarchically. For simplicity, we hereafter consider 
the case that each branching point has $N$ branches. 

We say that $d(\alpha,\beta)$, the distance between two states 
$\alpha$ and $\beta$, is \( k \) if the lowest common ancestor of these two 
states is in the \( k \)-th layer counted from the bottom, e.g., 
in Fig.~1 $d(\alpha,\beta)=1$ and $d(\alpha,\gamma)=2$. 
It is assumed that overlaps between states depend only on the distance 
and they decrease with increasing distance. 
Therefore $q_k$, the overlap between two states with $d(\alpha,\beta)=k$, 
satisfies the relation
\begin{equation}
q_0 > q_1 > \cdots > q_{L-1} > q_L=0.
\end{equation}

We hereafter denote as \( h_k(B_{k-1}) \) the barrier height from a 
branching point in the $k-1$-th layer, $B_{k-1}$, to its parent branching 
point. In the GREM, \( h_k \) is 
given randomly and independently according to the distribution
\begin{equation}
\rho_k(h_k){\rm d}h_k = \frac{1}{T_{\rm c}(k)}\exp[-\frac{h_k}{T_{\rm c}(k)}]
{\rm d}h_k,
\label{eqn:rhoE}
\end{equation}
where \( T_{\rm c}(k) \) is the transition temperature of the \( k \)-th layer 
and satisfies the inequality
\begin{equation}
T_{\rm c}(1) < T_{\rm c}(2) < \cdots < T_{\rm c}(L).
\label{eqn:Tcrelation}
\end{equation}

For dynamics, we employ Markoff process similar to that 
used in ref.~\citen{Bouchaud}. 
At first, the system is activated from a state 
\( \beta \) to \( \beta_k \) (the $k$-th {\em ancestor} of \( \beta \)) 
with probability \( W(\beta,k) \) 
in unit time. After the activation, the system falls 
to one of all the states under \( \beta_k \) with equal probability. 
We choose the uniform distribution for the initial condition to consider the 
situation that the system is quenched from an infinitely high temperature. 

The difference between the present dynamics and the one 
by Bouchaud and Dean exists in the hopping probability 
\( W(\beta,k) \). Inspired by the model proposed by 
Yoshino\cite{Yoshino}, we give \( W(\beta,k) \) as
\begin{equation}
W(\beta,k)=\frac{1}{\tau_0\prod_{i=1}^{k}{\tilde \tau}_i(\beta_{i-1})}
\left[1-\frac{1}{{\tilde \tau}_{k+1}(\beta_k)}\right],
\label{eqn:defofWk}
\end{equation}
where \( \tau_0 \) is a microscopic time scale, 
\( {\tilde \tau}_k \equiv \exp(h_k/T) \) and 
${\tilde \tau}_{L+1}(\beta_L)\equiv\infty$. 
Note that the first factor in the right hand represents 
the probability that the system can be activated from \( \beta \) 
to \( \beta_k \) and the second one represents that the system can not be 
activated from \( \beta_k \) to \( \beta_{k+1} \). 
Hereafter \( \tau_0 \) is used as the unit of time and set to \( 1 \). 
From eq.~(\ref{eqn:rhoE}), the distribution of \( {\tilde \tau_k} \) 
can be written as
\begin{equation}
p_k({\tilde \tau_k}) {\rm d} {\tilde \tau}_k = \frac {x_k}
{{\tilde \tau}_k^{x_k+1}}{\rm d} {\tilde \tau}_k 
\hspace{1cm}({\tilde \tau}_k \geq 1),
\label{eqn:sinpleptau}
\end{equation}
where \( x_k \equiv T/T_{\rm c}(k) \). 

On the other hand, Bouchaud and Dean give the hopping probability as
\begin{equation}
W(\beta,k)=\frac{1}{\tau_0 {\tilde \tau}_k(\beta_{k-1})}.
\label{eqn:WbyBouchaud}
\end{equation}
But this hopping probability is not so realistic because, e.g., 
$W(\beta,k+1)>W(\beta,k)$ if $h_{k+1}(\beta_k)<h_{k}(\beta_{k-1})$. 

Now let us introduce the following two functions. The first one 
,$\Pi(k,\alpha,t)$, is the probability that the system which initially 
stays at \( \alpha \) reaches a state with the distance \( k \) at 
\( t \). This function is defined by the Green function $G_{\beta\alpha}(t)$, 
i.e., the probability that the system which 
initially stays at \( \alpha \) reaches \( \beta \) at \(t\), as
\begin{equation}
\Pi(k,\alpha,t)\equiv\sum_{\beta:d(\alpha,\beta)=k} G_{\beta\alpha}(t),
\label{eqn:defofPi}
\end{equation}
where the sum of this equation is 
taken over all the states \( \beta \) which satisfy the condition 
$d(\alpha,\beta)=k$. 
The next function, \( P(\alpha,B,t_{\rm w}) \), is the probability that 
we find the system at a state \( \alpha \) at time \( t_{\rm w} \) 
when the initial condition is located anywhere under \( B \), and is defined 
by the Green function as
\begin{equation}
P(\alpha,B,t_{\rm w})=\frac{1}{N(B)}\sum_{\beta \in B}G_{\alpha\beta}
(t_{\rm w}),
\label{eqn:defofPB}
\end{equation}
where $N(B)$ is the number of states under $B$. 
Using these two functions, we can represent \(C(t+t_{\rm w},t_{\rm w})\), 
the correlation between \( t_{\rm w} \) and \( t+t_{\rm w} \), as
\begin{equation}
C(t+t_{\rm w},t_{\rm w})=\sum_{k=0}^{L-1}q_k\sum_{\alpha}\Pi(k,\alpha,t)
P(\alpha,B_{\rm top},t_{\rm w}).
\end{equation}
We hereafter calculate the correlation function along the following 
strategy:
\begin{itemize}
\item[(i)] Calculate the Laplace transform of 
\( P(\alpha,B_{\rm top},t_{\rm w}) \) 
\begin{equation}
{\hat P}(\alpha,B_{\rm top},E)\equiv \int_0^{\infty} {\rm d}t_{\rm w} 
\exp[-Et_{\rm w}] P(\alpha,B_{\rm top},t_{\rm w}).
\label{eqn:step1}
\end{equation}
\item[(ii)] Calculate the Green function and estimate
\begin{equation}
{\hat \Pi}(k,\alpha,E')\equiv\int_0^{\infty}{\rm d}t
\Pi(k,\alpha,t) \exp[-E't].
\label{eqn:step2}
\end{equation}
\item[(iii)] By taking the sum for \( \alpha \), calculate
\begin{equation}
{\hat C}(E'+E,E)\equiv \sum_{k=0}^{L-1}q_k \sum_{\alpha}
{\hat \Pi}(k,\alpha,E'){\hat P}(\alpha,B_{\rm top},E).
\label{eqn:step3}
\end{equation}
\item[(iv)] Carry out the inverse Laplace transformation of 
${\hat C}(E'+E,E)$ to obtain $C(t+t_{\rm w},t_{\rm w})$.
\end{itemize}
\section{Analytic Calculations and Simulations on the GREM}
\label{sec:analysis}
\subsection{Derivation of the formal solution}
Now let us consider the event that the system initially stays at \( \beta \) 
and reaches \( \alpha \) at time \( t_{\rm w} \). The probability for this event is 
\( P(\alpha,\beta,t_{\rm w})=G_{\alpha\beta}(t_{\rm w}) \). 
There are two possibilities for this event to happen: 
\begin{itemize}
\item[(i)] \( \alpha=\beta \) and the system is never activated 
during \( t_{\rm w} \). 
\item[(ii)] The system is activated 
to \( \beta_k \) at \( t' \) \( (<t_{\rm w}) \) and 
reaches \( \alpha \) after that. 
\end{itemize}
In the case (ii), the probability that the system reaches \( \alpha \) 
after the activation is \(  P(\alpha,\beta_k,t_{\rm w}-t') \) because the 
next state is chosen randomly from all the states under 
\( \beta_k \) (recall the definition of \( P(\alpha,B,t_{\rm w}) \)). 
Therefore, we obtain the following integral equation
\begin{eqnarray}
P(\alpha,\beta,t_{\rm w})
=\delta_{\alpha \beta}\exp[-\frac{t_{\rm w}}{{\tilde \tau}_1(\beta)}]
+\sum_{i=1}^{L}\int_0^{t_{\rm w}}{\rm d}t' W(\beta,i)
\exp[-\frac{t'}{{\tilde \tau}_1(\beta)}]P(\alpha,\beta_i,t_{\rm w}-t').
\end{eqnarray}
The Laplace transformation of this equation leads us to 
\begin{eqnarray}
{\hat P}(\alpha,\beta,E)=
\frac{\delta_{\alpha\beta}{\tilde \tau}_1(\beta)}{E{\tilde \tau}_1(\beta)+1}
+\sum_{i=1}^{L}\frac{W(\beta,i){\tilde \tau}_1(\beta)}
{E{\tilde \tau}_1(\beta)+1}
{\hat P}(\alpha,\beta_i,E).
\label{eqn:basiceq}
\end{eqnarray}

If we substitute the Laplace transformation of eq.~(\ref{eqn:defofPB}) into 
eq.~(\ref{eqn:basiceq}), it becomes linear equations of 
the Green functions, so that we can calculate 
the Green function \( {\hat P}(\alpha,\beta,E) \) itself and 
\( {\hat P}(\alpha,B_{\rm top},E) \). The calculation is described in 
appendix~\ref{sec:formalsolution} and the results are
\begin{equation}
{\hat P}(\alpha,B_{\rm top},E)
= \frac{{\tilde \tau}_1(\alpha)}{N^L \{E{\tilde \tau}_1(\alpha)+1\}
\prod_{l=1}^{L}Z_l(\alpha_l)},
\label{eqn:solofPEtop}
\end{equation}
and
\begin{equation}
{\hat G}_{\alpha \beta}(E)=\frac{{\tilde \tau}_1(\beta)\delta_{\alpha\beta}}
{E{\tilde \tau}_1(\beta)+1}+\sum_{k=1}^{L}\frac{1}{N^k}
\frac{{\tilde \tau}_1(\alpha){\tilde \tau}_1(\beta)W(\beta,k)
\delta_{\alpha_k\beta_k}}
{\{E{\tilde \tau}_1(\alpha)+1\}\{E{\tilde \tau}_1(\beta)+1\}
Z_k(\beta_k)\prod_{l=1}^{k-1}Z_l(\alpha_l)Z_l(\beta_l)},
\label{eqn:solofGreen}
\end{equation}
where the function \( Z_k \) is defined as
\begin{equation}
Z_{k}(B_{k})\equiv1-[1-\{{\tilde \tau}_{k+1}(B_{k})\}^{-1}]X_{k}(B_{k})
\hspace{3mm}(0\le k \le L),
\label{eqn:Zfork}
\end{equation}
and \( X_k \) is defined by the following recursive equations,
\begin{subeqnarray}
X_0(B_0)&\equiv& \frac{1}{E+1},\\
X_{k+1}(B_{k+1})&\equiv& \frac{1}{N}\sum_{B_{k}\in B_{k+1}}\frac{X_{k}(B_{k})}
{\{1-X_{k}(B_{k})\}{\tilde \tau}_{k+1}(B_{k})+X_{k}(B_{k})}.
\label{eqn:Xfork}
\end{subeqnarray}
\subsection{Calculations of the correlation function}
Let us calculate \( X_k(B_k) \) in the explicit form. 
From eq.~(\ref{eqn:Xfork}), we can expect that 
\( X_k(B_k) \) does not depend on \( B_k \) in the limit 
\( N\rightarrow \infty \) (we hereafter simply denote as \( X_k \)) 
and is calculated recursively as
\begin{equation}
X_k=x_k I(V_k,x_k)\hspace{1cm}(1\le k \le L),
\end{equation}
\begin{equation}
I(V,x)\equiv \int_1^{\infty }\frac{{\rm d} u}{u}
\frac{u^{-x}}{Vu+1},
\end{equation}
where
\begin{equation}
V_k = \frac{1-X_{k-1}}{X_{k-1}}.
\label{eqn:REforAk}
\end{equation}
In the following calculations, we consider the case \( x_L<\cdots<x_1<1 \). 
In the assumption \( E \ll 1 \) (equivalent to $t_{\rm w} \gg 1$ ), we find
\begin{equation}
X_k\approx 1-C_kE^{\gamma_k}\hspace{3mm}(1\le k \le L),
\end{equation}
where
\begin{subeqnarray}
\gamma_k &=& \prod_{l=1}^k x_l,\\
C_k&\equiv&\left\{ 
  \begin{array}{cl}
\Gamma(1-x_1)\Gamma(1+x_1)
&\mbox{($ k=1 $)}, \vspace{2mm}\\
\Gamma(1-x_k)\Gamma(1+x_k)C_{k-1}^{x_k}
&\mbox{($ k\ge 2 $)}.\hspace{1cm}
\end{array}\right.
\label{eqn:XCfork}
\end{subeqnarray}
By using this result to eqs.(\ref{eqn:Zfork}) and~(\ref{eqn:solofPEtop}), 
${\hat P}(\alpha,B_{\rm top},E)$ is explicitly given as
\begin{equation}
{\hat P}(\alpha,B_{\rm top},E)\approx
\frac{E^{-\gamma_L}\prod_{m=1}^L{\tilde \tau}_m(\alpha_{m-1})}
{C_L N^L\{E{\tilde \tau}_1(\alpha)+1\}
\prod_{l=1}^{L-1}[C_l E^{\gamma_l}{\tilde \tau}_{l+1}(\alpha_l) +1]},
\label{eqn:finalPBtop}
\end{equation}
where we have used the fact \( {\tilde \tau}_{L+1}=\infty \). 

Next we estimate \( {\hat \Pi}(k,\alpha,E')\) defined by 
eqs.~(\ref{eqn:defofPi}) and~(\ref{eqn:step2}). 
By taking the leading order of $\frac1N$ in eq.~(\ref{eqn:solofGreen}), we find
\begin{subequations}
\begin{equation}
{\hat \Pi}(0,\alpha,E')\approx\frac{{\tilde \tau}_1(\alpha)}
{E'{\tilde \tau}_1(\alpha)+1},
\end{equation}
\begin{eqnarray}
{\hat \Pi}(k,\alpha,E')&\approx&\frac{{\tilde \tau}_1(\alpha)W(\alpha,k)}
{{\{E'{\tilde \tau}_1(\alpha)+1\}\prod_{l=1}^{k}Z_l(\alpha_l)}}
\frac{1}{N^k}\sum_{\beta:d(\alpha,\beta)=k}
\frac{{\tilde \tau}_1(\beta)}{\{E'{\tilde \tau}_1(\beta)+1\}
\prod_{l=1}^{k-1}Z_l(\beta_l)}\nonumber\\
&\approx&\frac{{\tilde \tau}_1(\alpha)W(\alpha,k)}
{{\{E'{\tilde \tau}_1(\alpha)+1\}\prod_{l=1}^{k}Z_l(\alpha_l)}}
 x_1I(E',x_1-1)\prod_{l=1}^{k-1}x_{l+1}
I(C_l E'^{\gamma_l},x_{l+1}-1)\nonumber\\
&=&\frac{C_k {\tilde \tau}_1(\alpha)W(\alpha,k)
E'^{\gamma_k-1}}
{{\{E'{\tilde \tau}_1(\alpha)+1\}\prod_{l=1}^{k}Z_l(\alpha_l)}}
\hspace{3mm}(1\le k \le L).
\end{eqnarray}
\end{subequations}

We are now in the position to calculate eq.~(\ref{eqn:step3}). 
At first, let us define a function
\begin{equation}
{\hat C}_k(E'+E,E)\equiv \sum_{\alpha} {\hat P}(\alpha,B_{\rm top},E)
{\hat \Pi}(k,\alpha,E').
\end{equation}
For \( k=0 \), this function is estimated as
\begin{eqnarray}
{\hat C}_0(E'+E,E) &\approx& \frac{E^{-\gamma_L}}{C_L}x_1J(E,E';x_1-2)
\prod_{l=1}^{L-1}x_{l+1}I(C_l E^{\gamma_l},x_{l+1}-1)\nonumber\\
&=&{\hat g}(E',E;\gamma_1)+\frac{1}{E(E'-E)},
\label{eqn:lastC0}
\end{eqnarray}
where
\begin{equation}
J(A,B;\nu)\equiv\int_1^{\infty}\frac{{\rm d}s}{s}\frac{s^{-\nu}}{(As+1)(Bs+1)},
\end{equation}
and
\begin{equation}
{\hat g}(E',E;\alpha)\equiv\frac{E'^{\alpha-1}E^{-\alpha} }{E-E'}.
\end{equation}
Similarly, for \( k \ge 1 \)
\begin{eqnarray}
{\hat C}_k(E'+E,E)&\approx&
\frac{E^{-\gamma_L} E'^{\gamma_k-1} C_k}{C_L}x_1 J\left(E,E';x_1-1\right)
\prod_{l=1}^{k-1}x_{l+1}J\left(C_l E^{\gamma_l},C_l E'^{\gamma_l}
;x_{l+1}-1\right)
\nonumber\\
&&\times
x_{k+1}\Bigl\{ J\left(C_k E^{\gamma_k},C_k E'^{\gamma_k};x_{k+1}-2\right)
-J\left(C_k E^{\gamma_k},C_k E'^{\gamma_k};x_{k+1}-1\right)\Bigr\}\nonumber\\
&&\times
\prod_{m=k+1}^{L-1}x_{m+1} I(C_m E^{\gamma_m},x_{m+1}-1)
\nonumber\\
&\approx& g(E',E;\gamma_{k+1})-g(E',E;\gamma_{k}),
\label{eqn:lastCk}
\end{eqnarray}
where we have used the fact 
$J\left(C_k E^{\gamma_k},C_k E'^{\gamma_k};x_{k+1}-1\right)$ is negligible to 
$J\left(C_k E^{\gamma_k},C_k E'^{\gamma_k};x_{k+1}-2\right)$.

Now let us carry out the inverse Laplace transformation of 
eqs.(\ref{eqn:lastC0}) and~(\ref{eqn:lastCk}). At first, 
\( \frac{1}{E(E'-E)} \) is transformed to \( 1 \). 
As for \( {\hat g}(E',E;\alpha) \), the transformation from 
\( E' \) to \( t \) leads us to
\begin{equation}
g(t,E;\alpha)=\frac{-1}{\Gamma(1-\alpha)}
\int_0^t {\rm d}s (t-s)^{-\alpha}E^{-\alpha}{\rm e}^{Es}.
\end{equation}
Then by transforming from \( E \) to \( t_{\rm w} \), we find
\begin{eqnarray}
g(t,t_{\rm w};\alpha) = f\left(\frac{t}{t+t_{\rm w}};\alpha\right)-1,
\label{eqn:IRTforg}
\end{eqnarray}
where
\begin{equation}
f(x;\alpha)\equiv\frac{1}{\Gamma(\alpha)\Gamma(1-\alpha)}\int_{x}^1
{\rm d}u(1-u)^{\alpha-1}u^{-\alpha}.
\end{equation}
From these results, we finally obtain
\begin{eqnarray}
C(t+t_{\rm w},t_{\rm w})&=&\sum_{k=0}^{L-1}q_k C_k(t+t_{\rm w},t_{\rm w})
\nonumber\\
&=& q_0 f\left(\frac{t}{t+t_{\rm w}};\gamma_{1}\right)
+\sum_{k=1}^{L-1}q_k \left\{
f\left(\frac{t}{t+t_{\rm w}};\gamma_{k+1}\right)
-f\left(\frac{t}{t+t_{\rm w}};\gamma_{k}\right)
\right\}.
\label{eqn:finalresult}
\end{eqnarray}
The asymptotic behavior in regimes \( t\ll t_{\rm w} \) and 
\( t \gg t_{\rm w} \) is
\begin{subeqnarray}
C_0(t+t_{\rm w},t_{\rm w})&\sim&\left\{ 
  \begin{array}{cl}
\displaystyle{1-(t/t_{\rm w})^{1-\gamma_1}}
&\mbox{($ t\ll t_{\rm w} $)} \vspace{6mm},\\
\displaystyle{(t/t_{\rm w})^{-\gamma_1}}
&\mbox{($ t\gg t_{\rm w} $)},
\end{array}\right.\\
&&\nonumber\\
C_k(t+t_{\rm w},t_{\rm w})&\sim&\left\{ 
  \begin{array}{cl}
\displaystyle{(t/t_{\rm w})^{1-\gamma_k}}
&\mbox{($ t\ll t_{\rm w} $)} \vspace{6mm},\\
\displaystyle{(t/t_{\rm w})^{-\gamma_{k+1}}}
&\mbox{($ t\gg t_{\rm w} $)}\hspace{1cm}(1\le k\le L-1).
\end{array}\right.
\label{eqn:asymptoC_k}
\end{subeqnarray}
As a result,
\vspace{2mm}
\begin{equation}
C(t+t_{\rm w},t_{\rm w})\sim \left\{ 
  \begin{array}{cl}
\displaystyle{q_0-(q_0-q_1)(t/t_{\rm w})^{1-\gamma_1}}
&\mbox{($ t\ll t_{\rm w} $)}, \vspace{2mm}\\
\displaystyle{q_{L-1}(t/t_{\rm w})^{-\gamma_L}}
&\mbox{($ t\gg t_{\rm w} $)}.\hspace{1cm}
\end{array}\right.
\vspace{2mm}
\label{eqn:asymptoC}
\end{equation}
\subsection{Calculations of the ZFC magnetization}
Now let us explain how to introduce magnetizations to the 
GREM\cite{condmat}. We assign the value of magnetization to 
a state \( \alpha \) as
\begin{equation}
M_{\alpha} ={\cal M}_0(\alpha)+{\cal M}_1(\alpha_1)+\cdots
+{\cal M}_{L-1}(\alpha_{L-1}),
\end{equation}
where \( {\cal M}_k(\alpha_k) \) is a contribution from the branching 
point $\alpha_k$, and is given from a distribution function 
\( {\cal D}_k ({\cal M}_k) \) as an independent random variable 
with zero mean. 
Note that ${\cal M}_n (n=k,k+1,\cdots)$ are common between two states 
\( \alpha \) and \( \beta \) if \( d(\alpha,\beta)=k \). This means that 
there is a correlation of magnetizations between two states, 
whose averaged value \( \overline{M_{\alpha}M_{\beta}} \) is
\begin{equation}
\overline{M_{\alpha}M_{\beta}}=\sum_{n=k}^{L-1} \overline{{\cal M}_n^2}
\hspace{1cm}(d(\alpha,\beta)=k),
\end{equation}
where \( \overline{{\cal M}_n^2} \) denotes the variance of 
\( {\cal D}_n ({\cal M}_n) \). Therefore we can obtain the correlation 
function of magnetization by setting \( q_k \) in eq.~(\ref{eqn:finalresult}) 
to \( \sum_{n=k}^{L-1} \overline{{\cal M}_n^2} \). 

The effect of magnetic field \( H \) is considered by changing 
\( h_1(\beta) \) into \( h_1(\beta)+HM_{\beta} \). As a result, 
the hopping rate \( W(\beta,k) \) and the transition probability 
from \( \beta \) to any states are modified by a factor 
\( \exp(-H M_{\beta}) \). Therefore, as mentioned in 
\( \S \)~\ref{sec:introduction}, we can use the relation 
eq.~(\ref{eqn:relationRC}) to calculate \( M_{\rm ZFC}(t,t_{\rm w}) \) as
\begin{eqnarray}
M_{\rm ZFC}(t,t_{\rm w})&\equiv& H\int_{t_{\rm w}}^{t+t_{\rm w}} 
{\rm d} t' R(t+t_{\rm w},t') \nonumber \\
&=& \sum_{k=0}^{L-1} \frac{\beta H \overline{{\cal M}_k^2}}
{\Gamma(\gamma_k)\Gamma(1-\gamma_k)}\int_{\frac{t_{\rm w}}{t+t_{\rm w}}}^1
{\rm d}u u^{\gamma_k}(1-u)^{-\gamma_k}.
\label{eqn:ZFCresult}
\end{eqnarray}
\subsection{Simulations}
To check the validity of analytic results obtained above, 
we carry out simulations on the GREM 
with \( L=3 \), \( x_1=0.6, x_2=0.4 \) and \( x_3=0.3 \) 
( see ref.~\citen{condmat} for the details of the simulation). 
We first observe \( C_k(t+t_{\rm w},t_{\rm w}) \), 
the probability that the distance between the states at 
\( t_{\rm w} \) and \( t+ t_{\rm w} \) is \( k \), 
for \( t_{\rm w}=10^2,10^3,\ldots,10^7 \). 
We take a random average over \( 10^6 \) samples. In Fig.~2, we plot 
\( C_k(t+t_{\rm w},t_{\rm w}) \) as a function of \( t/t_{\rm w} \) with 
the analytic result shown in eq.~(\ref{eqn:finalresult}). 
We can see that the data are rather consistent with the analytic result. 

Next we observe the ZFC magnetization \( M_{\rm ZFC}(t,t_{\rm w}) \) 
in the case that \( H/T_{\rm c}(3)=0.1 \) and 
\( \overline{{\cal M}_k^2} =\frac19 \) (\(k=0,1,2 \)) for 
\( t_{\rm w}=10^2,3\times10^2,10^3,\ldots,10^5 \). The number 
of samples used for a random average is \(10^8 \). In Fig.~3, 
the scaling plot of \( M_{\rm ZFC}(t,t_{\rm w}) \) is shown 
with the analytic result, eq.~(\ref{eqn:ZFCresult}). The validity of the 
analytic result is well confirmed. 
\section{Summary and Discussions}
In this manuscript, we have introduced more realistic dynamics 
than that used in ref.~\citen{Bouchaud} to the GREM, and calculated 
the time correlation function exactly. The results obtained by 
Bouchaud and Dean and ours are very similar in the sense that 
\( t/t_{\rm w} \) scaling holds in both calculations and 
the asymptotic behavior shown in eqs.~(\ref{eqn:asymptoC_k}) 
and (\ref{eqn:asymptoC}) is the same. 
But in ref.~\citen{Bouchaud}, all the exponents \(\gamma_k\) ($k=1,\ldots,L$) 
in eqs.~(\ref{eqn:asymptoC_k}) and (\ref{eqn:asymptoC})
are replaced with \( x_k \). The reason for 
this difference on the exponents is that, as seen from 
eqs.~(\ref{eqn:defofWk}) and~(\ref{eqn:WbyBouchaud}),
the hopping rate \( W(\beta,k) \) depends on 
\( {\tilde \tau}_i(\beta) \) ($i=1,\ldots,k$) in our calculation, 
but only on \( {\tilde \tau}_k(\beta) \) in ref.~\citen{Bouchaud}. 

We have also calculated the ZFC magnetization with the 
relation eq.~(\ref{eqn:relationRC}) and shown the validity by simulation. 
From this result, we have confirmed that eq.~(\ref{eqn:relationRC}) is 
really valid in the GREM. But, as mentioned in 
\( \S \)~\ref{sec:introduction}, this relation holds only on the 
specific models and kinetics. Actually, Cugliandolo and Kurchan~\cite{Kurchan} 
proposed a different relation between the correlation function and the 
response function on the SK model and confirmed it numerically. 
It is very interesting subject to investigate how these two functions 
are related in other systems in non-equilibrium.

Now let us comment on two main reasons which seem to make this model 
solvable. The first one is that the number of branches in each layer 
is very large. For example, as seen from 
eqs.~(\ref{eqn:solofPEtop}), (\ref{eqn:Zfork}) and~(\ref{eqn:Xfork}), 
\( {\hat P}(\alpha,B_{\rm top},E) \)
depends on the full structure of the tree (\( {\tilde \tau}_k \) of 
all the branches, to be specific). But this quantity is self-averaging and 
only \( {\tilde \tau}_k(\alpha_{k-1}) \) (\(k=1,2,\ldots,L\)) dependence 
remains in the limit \( N\rightarrow \infty \). The second one is the 
choice of the hopping rate \( W(\beta,k) \). From eq.~(\ref{eqn:finalPBtop}), 
we can see that 
\( {\hat P}(\alpha,B_{\rm top},E) \) depends on 
\( {\tilde \tau}_k(\alpha_{k-1}) \) as the following form
\begin{equation}
{\hat P}(\alpha,B_{\rm top},E)
=\prod_{k=1}^{L} F_k\{E,{\tilde \tau}_k(\alpha_{k-1})\}.
\end{equation}
This means that ${\hat P}(\alpha,B_{\rm top},E)$ depends on 
\( {\tilde \tau}_k(\alpha_{k-1}) \) independently. As seen above, 
this property makes our calculation simple. We confirmed that 
this property relies on the choice of \( W(\beta,k) \) and does not hold 
if, for example, we set it as
\begin{equation}
W(\beta,k)=\frac{1}{\prod_{i=1}^{k}{\tilde \tau}_i(\beta_{i-1})}.
\end{equation}

In this manuscript, we have succeeded to calculate the Green function of 
this model. By using it directly, we may be able to carry out simulations 
as done in ref.~\citen{condmat} in longer time regime. 
\section*{Acknowledgments}
We would like to thank H. Yoshino for valuable discussions and
suggestions on the manuscript. The numerical calculations were performed 
on an Origin 2000 at Division of Physics, Graduate school of Science, 
Hokkaido University. 
\newpage
\appendix
\section{Derivation of eqs.~(\ref{eqn:solofPEtop}) and (\ref{eqn:solofGreen})}
\label{sec:formalsolution}
We first prove the following equations:
\begin{subeqnarray}
{\hat P}(\alpha,B_k,E)&=&\frac{Y_k(\alpha)\delta_{B_k\alpha_k}}{Z_k(B_k)}
+\sum_{i=k+1}^{L}b(B_k,i){\hat P}(\alpha,A_i(B_k),E)
\hspace{3mm}(0\le k \le L),\\
b(B_k,i)&=&\frac{[1-\{{\tilde \tau}_{i+1}[A_i(B_k)]\}^{-1}]X_k(B_k)}
{Z_k(B_k) \prod_{l=k+1}^{i}{\tilde \tau}_{l}[A_{l-1}(B_k)]},
\label{eqn:hatP}
\end{subeqnarray}
where \( \alpha_0=\alpha \). The notation 
\( A_m(B_n) \) (\( m \ge n \)) stands for the branching point which is 
the ancestor of \( B_n \) and is in the \( m \)-th layer. The functions 
\( X_k \) and \( Z_k \) are defined by eqs.~(\ref{eqn:Xfork}) and 
(\ref{eqn:Zfork}), and \( Y_k \) is defined as
\begin{subeqnarray}
Y_0(B_0)&\equiv&\frac{1}{E+1},\\
Y_{k}(\alpha)&\equiv&\frac{Y_{0}(\alpha)}{N^k\prod_{l=0}^{k-1}Z_l(\alpha_l)}
\hspace{3mm}(1\le k \le L).
\label{eqn:Yfork}
\end{subeqnarray}
We can easily check that eq.~(\ref{eqn:hatP}) is reduced to 
eq.~(\ref{eqn:basiceq}) for \( k=0 \). 
Next we can prove eq.~(\ref{eqn:hatP}) in terms of mathematical induction 
by substituting eq.~(\ref{eqn:hatP}) for \( k=n \) into the relation
\begin{equation}
{\hat P}(\alpha,B_{n+1},E)=\frac{1}{N}\sum_{B_{n}\in B_{n+1}}
{\hat P}(\alpha,B_{n},E)\hspace{1cm}(1\le n \le L-1).
\end{equation}
Equation.~(\ref{eqn:solofPEtop}) is derived from eq.~(\ref{eqn:hatP}) 
for \(k=L\).

For other branching points, \( {\hat P}(\alpha,B_k,E) \) is given as
\begin{equation}
{\hat P}(\alpha,B_k,E)=\sum_{i=k}^{L}a(B_k,i)Y_i(\alpha)
\delta_{A_i(B_k),\alpha_i}
\hspace{1cm}(0\le k\le L),
\label{eqn:RE2}
\end{equation}
\begin{equation}
a(B_k,i) = \left\{ 
  \begin{array}{cl}
\displaystyle{\frac{1}{Z_k(B_k)}} &\mbox{($ i=k $)}, \vspace{2mm}\\
\displaystyle{\frac{X_k(B_k)[1-\{{\tilde \tau}_{i+1}[A_i(B_k)]\}^{-1}]}
{Z_k(B_k)\prod_{l=k+1}^{i}Z_l[A_l(B_k)]{\tilde \tau}_{l}[A_{l-1}(B_k)]}}
&\mbox{($ i>k $)}.
\end{array}\right.
\vspace{5mm}
\label{eqn:DEFa}
\end{equation}
From these equations for \( k=0 \), we can derive eq.~(\ref{eqn:solofGreen}). 
We hereafter prove these equations recursively. For \( k=L \), they are 
obviously valid from eq.~(\ref{eqn:solofPEtop}). Next we show that 
eqs.~(\ref{eqn:RE2}) and~(\ref{eqn:DEFa}) are valid for \( k=n-1 \) 
if they hold for \( k\ge n \). By substituting eq.~(\ref{eqn:RE2}) 
into the right hand of eq.~(\ref{eqn:hatP}a) for \( k=n-1 \), we find
\begin{equation}
{\hat P}(\alpha,B_{n-1},E)=\frac{Y_{n-1}(\alpha)\delta_{B_{n-1}\alpha_{n-1}}}
{Z_{n-1}(B_{n-1})} +\sum_{i=n}^{L} S(B_{n-1},i)Y_i(\alpha)
\delta_{A_{i}(B_{n-1}),\alpha_{i}},
\end{equation}
where
\begin{equation}
S(B_{n-1},i)=\sum_{k=n}^i b(B_{n-1},k)a[A_k(B_{n-1}),i],
\label{eqn:defofS}
\end{equation}
Therefore eqs.~(\ref{eqn:RE2}) and~(\ref{eqn:DEFa}) are proved 
for \( k=n-1 \) if \( S(B_{n-1},i)=a(B_{n-1},i)\) \( (i\ge n) \). 
By substituting eqs.~(\ref{eqn:hatP}b) and~(\ref{eqn:DEFa}) into 
eq.~(\ref{eqn:defofS}), it is shown that
\begin{eqnarray}
S(B_{n-1},i)&=&\frac{X_{n-1}(B_{n-1})[1-\{{\tilde \tau}_{i+1}
[A_i(B_{n-1})]\}^{-1}]}
{ Z_{n-1}(B_{n-1})\prod_{m=n}^i {\tilde \tau}_m[A_{m-1}(B_{n-1})]}
\nonumber \\
&&\times\left\{\frac{1}{Z_i[A_i(B_{n-1})]}+\sum_{k=n}^{i-1}
\frac{1-Z_k[A_k(B_{n-1})]}
{ \prod_{l=k}^i Z_l[A_l(B_{n-1})]}
\right\}\nonumber\\
&=&a(B_{n-1},i),
\end{eqnarray}
where we have used eq.~(\ref{eqn:Zfork}).

\newpage
\noindent
{\bf \large FIGURE CAPTIONS}

\vspace*{3mm}\noindent
Fig.~1  The structure of the Generalized Random Energy Model with \( L=2 \) 
        and \( N=5 \). 

\vspace*{3mm}\noindent
Fig.~2  $C_k(t+t_{\rm w},t_{\rm w})$ versus \( t/t_{\rm w} \) for 
        \( t_{\rm w}=10^2,10^3,\ldots,10^7 \). The solid line indicates the 
        analytic result obtained in eq.~(\ref{eqn:finalresult}). 

\vspace*{3mm}\noindent
Fig.~3  $M_{\rm ZFC}(t,t_{\rm w})$ versus \( t/t_{\rm w} \) for 
        \( t_{\rm w}=10^3,3\times10^3,10^4,\ldots,10^6 \). The solid line 
        indicates the analytic result obtained in eq.~(\ref{eqn:ZFCresult}).

\newpage
\begin{center}
\epsfile{file=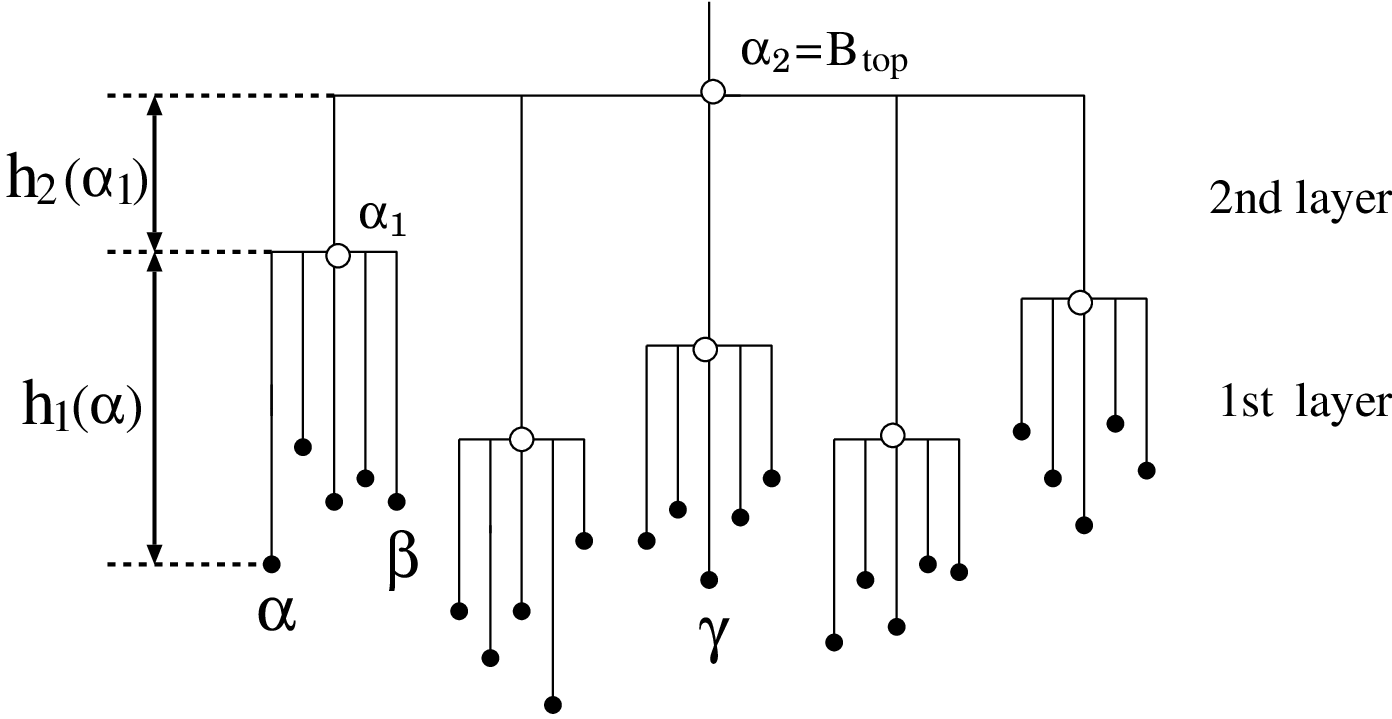,width=17cm}
\end{center}
\vspace{3cm}
\begin{center}
{\LARGE Fig.1}
\end{center}
\newpage
\begin{center}
\epsfile{file=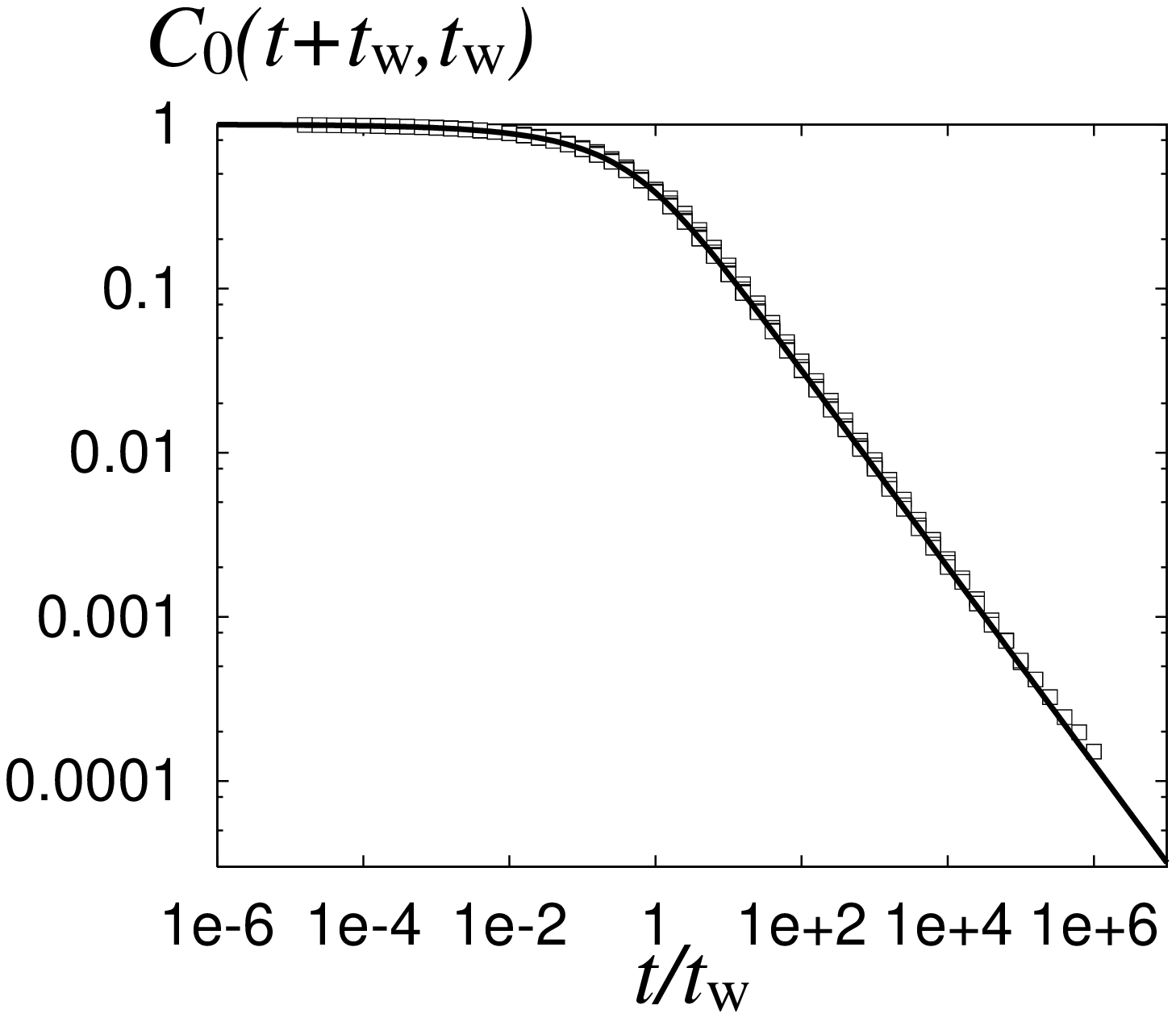,width=17cm}
\end{center}
\vspace{3cm}
\begin{center}
{\LARGE Fig.2(a)}
\end{center}
\newpage
\begin{center}
\epsfile{file=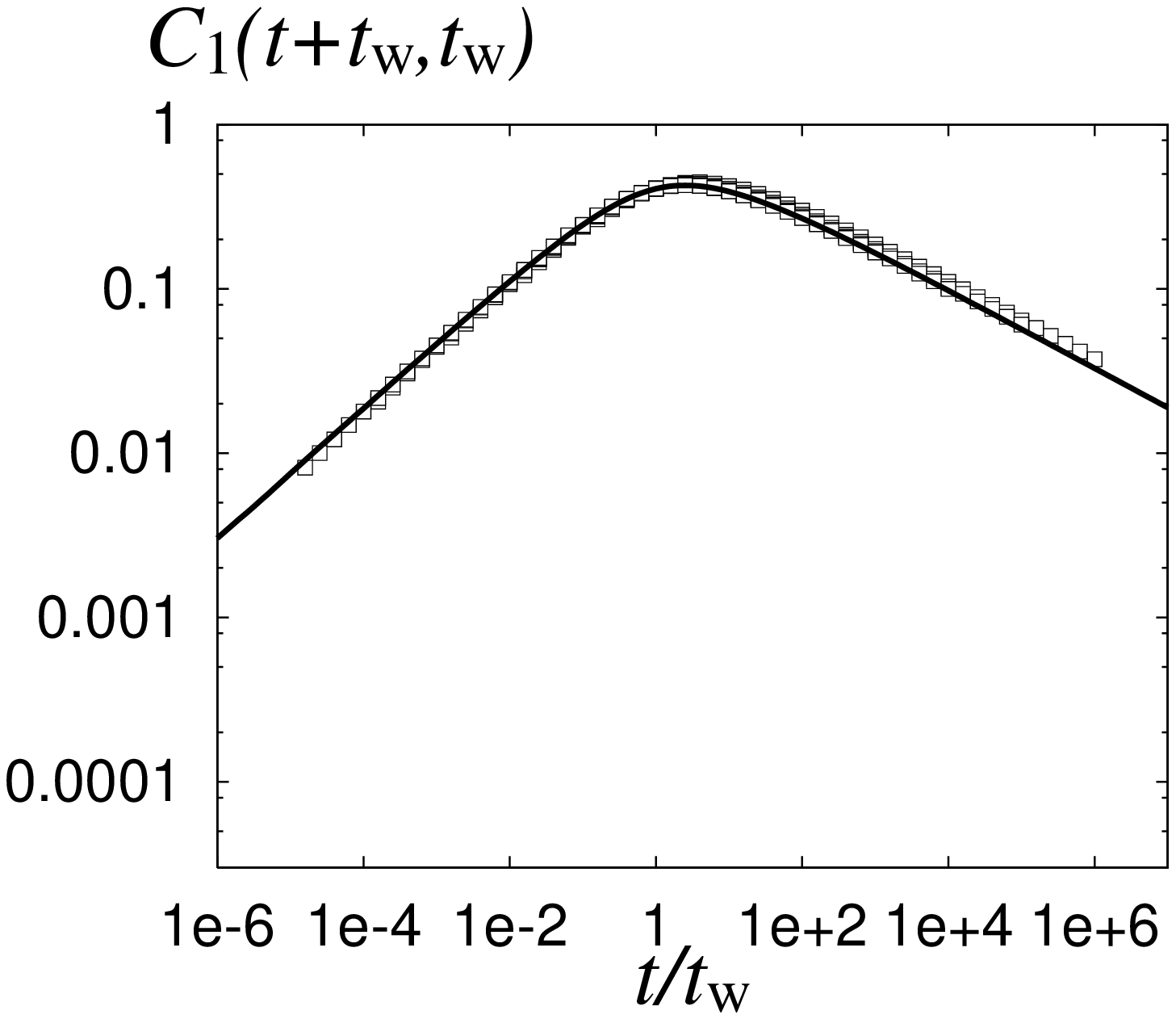,width=17cm}
\end{center}
\vspace{3cm}
\begin{center}
{\LARGE Fig.2(b)}
\end{center}
\newpage
\begin{center}
\epsfile{file=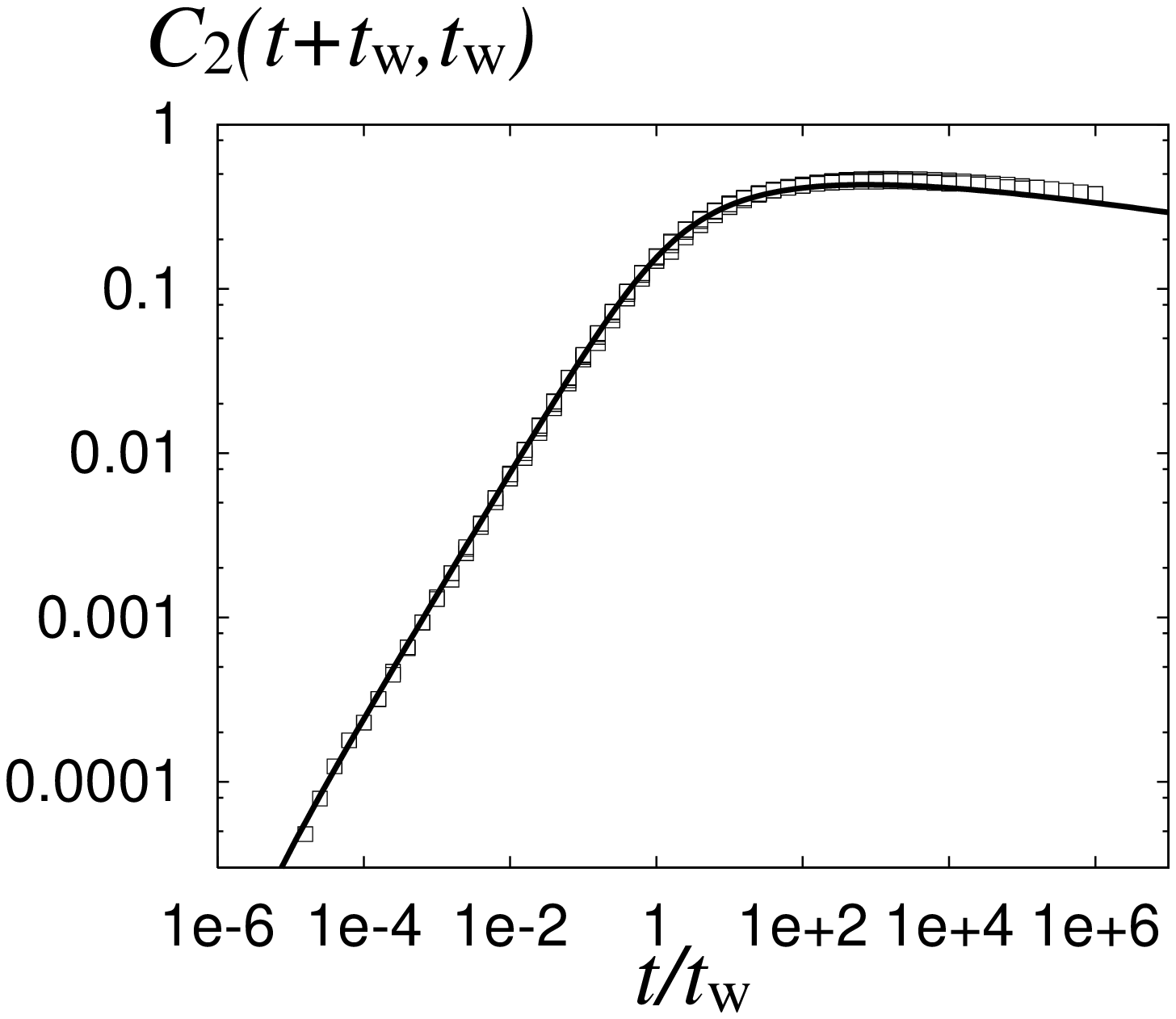,width=17cm}
\end{center}
\vspace{3cm}
\begin{center}
{\LARGE Fig.2(c)}
\end{center}
\newpage
\begin{center}
\epsfile{file=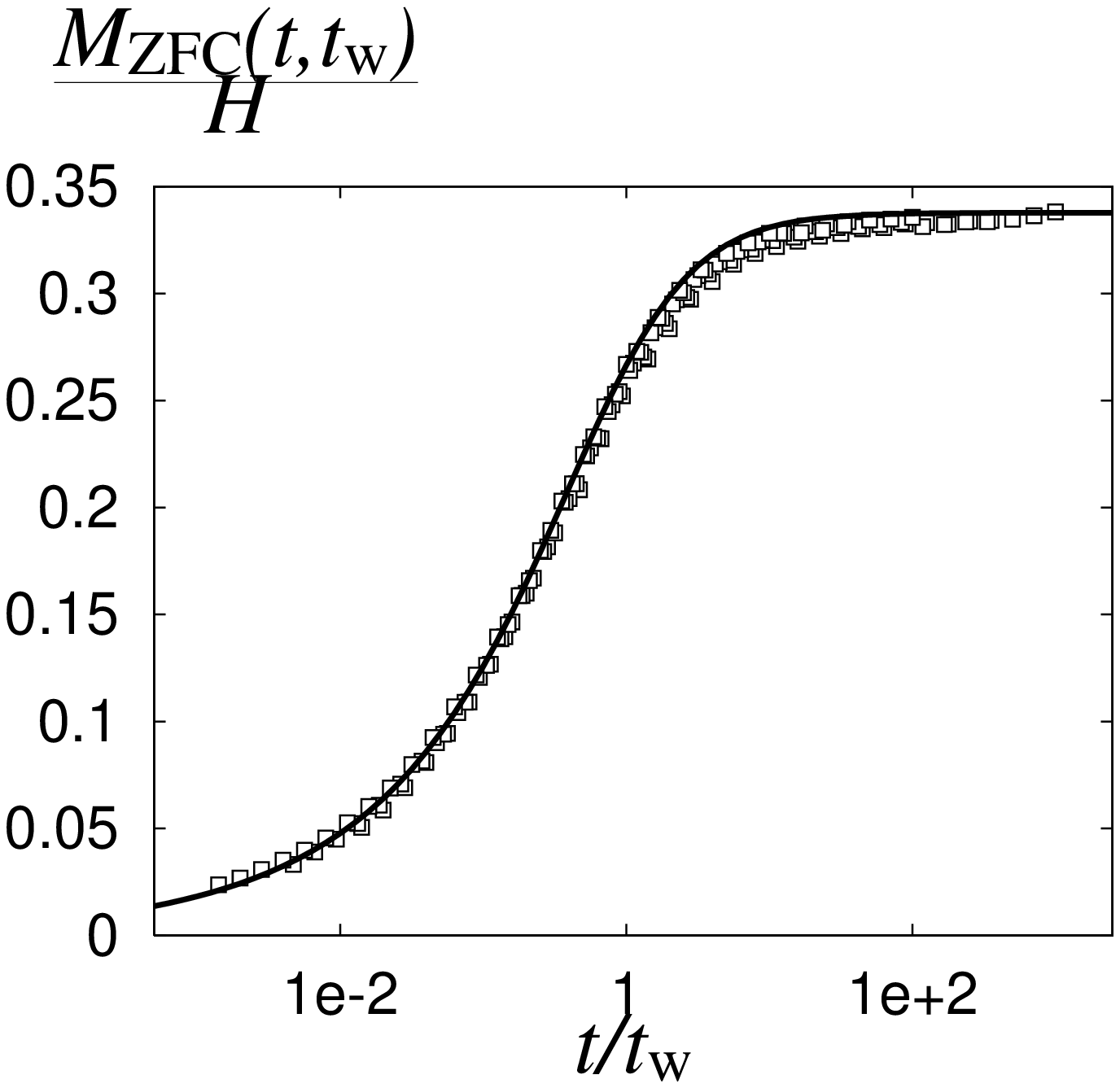,width=17cm}
\end{center}
\vspace{3cm}
\begin{center}
{\LARGE Fig.3}
\end{center}

\end{document}